\documentstyle[11pt,paspconf,epsf]{article}

\input epsf
\newdimen\fighsize \def\epsscale#1{\fighsize=#1\hsize} \epsscale{1}
\def\plotone#1{\epsfxsize=\fighsize\centerline{\epsfbox{#1}}}

\def\hub{$H_0$}

\begin{document}

\title{Pixellated Lenses and Estimates of $H_0$ from Time-delay Quasars}

\author{Liliya L.R. Williams}
\affil{Department of Physics and Astronomy, Univ. of Victoria, B.C, Canada}

\author{Prasenjit Saha}
\affil{Department of Physics, University of Oxford, UK}

\begin{abstract}
The largest source of uncertainty in the determination of \hub~from a
multiply-imaged QSO system is the unknown mass distribution in the
lensing galaxy. Parametric models severely restrict the shape of the
galaxy thereby underestimating errors and biasing the best estimate of
\hub. We present a method that explores the whole of the model space
allowed by the image observables and a few general properties of
galaxies. We describe blind tests of the method and then apply it
to PG1115+080 and B1608+656 which yield that \hub~ is between 45 and 
80 km~s$^{-1}$~Mpc$^{-1}$ at $90\%$ confidence level, with the best 
estimate of 60 km~s$^{-1}$~Mpc$^{-1}$.

\end{abstract}


\section{Introduction}

It has been 35 years since Refsdal (1964) proposed an elegant method to
derive the Hubble constant, \hub, from a multiply-imaged QSO system. 
Until a few years ago the main obstacle to implementing the method was the 
lack of sufficiently accurate data on the image positions and time delays 
between images. As the precision of the observational measurements improves 
the errors in \hub~ become dominated 
by the uncertainties in the galaxy mass distribution. These errors are very
hard to quantify using parametric shape(s) for the galaxy lens model; the
derived errors will tend to be underestimated as was noted by Bernstein and 
Fischer (1999) who constructed many types of parametric models for Q0957+561:
`The bounds on \hub~ are strongly dependent on our assumptions about a 
``reasonable'' galaxy profile.' In this contribution we develop and apply a
non-parametric method for modeling lensing galaxies and thus estimating \hub~
with the errorbars derived entirely from the uncertainties in the galaxy 
mass map. The method was initially applied to reconstructing mass maps of
lensing galaxies (Saha \& Williams 1997); in this volume, Saha et al. show
how it can be extended to recover mass distribution in galaxy clusters.

\section{The Method}

We start by tiling the lens plane with $\sim 25^2$ independent mass pixels 
each $\sim 0.1''$
on the side. Then we pixellate the lens equation and the arrival time surface.
The image observables, which we take to be exact, enter as primary modeling
constraints. The secondary constraints pertain to the main lensing galaxy:
\begin{enumerate}
\item mass pixel values, $\kappa_n$ must be non-negative;
\item location of the galaxy center is assumed to be coincident with that of
the optical/IR image;
\item the gradient of the projected mass density of the galaxy must point 
away from galaxy center to within a tolerance angle of $\pm 45^\circ$;
\item inversion symmetry, i.e. a galaxy must look the same if rotated by 
$180^\circ$ (applied only if galaxy optical/IR image looks symmetric);
\item logarithmic projected density gradient in the image region,
${{d~log\kappa}\over{d~logr}}={\rm ind}(r)$ should not be any shallower 
that -0.5;
\item external shear, i.e. influence of mass other than the main lensing 
galaxy is restricted to be constant across the image region, i.e. it is 
represented by adding a term
${1\over 2}\gamma_1(\theta_1^2-\theta_2^2)+\gamma_2(\theta_1\theta_2)$ to the
lensing potential.
\end{enumerate}
\begin{figure}
\epsscale{0.33}
\plotone{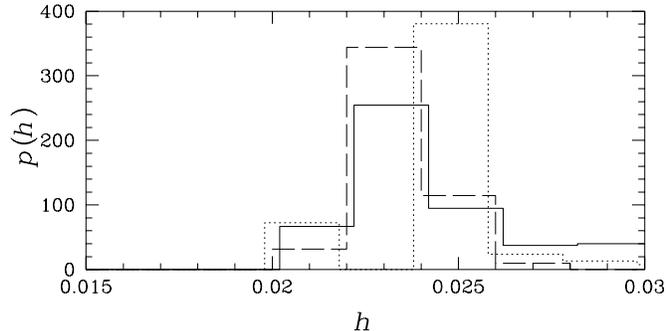}
\caption{Probability distribution of $h$ derived using four synthetic lens 
systems. Dashed line: inversion symmetry applied to all 4 galaxies, dotted 
line: inversion symmetry constraint not used, solid line: inversion symmetry
used where modeler deemed it appropriate.}\label{fig-1}
\end{figure}
After both primary and secondary constraints have been applied, we are 
left with a large number of viable galaxy models. We then generate a fair 
sample of the remaining model space by random walking through it until we 
accumulate 100 mass maps. Each of these galaxies reproduce the image 
properties exactly and each looks reasonably like a real galaxy. The 
distribution of the 100
corresponding \hub's is therefore the derived probability distribution
$p(h)$. The width of the distribution indicates the uncertainty arising
from our lack of sufficient knowledge about the lensing galaxy.
(See Williams \& Saha 1999 for details.)

\section{Blind tests}

First we apply the method to a synthetic sample of lenses. One of us 
constructed 4 sets of QSO-galaxy four-image lenses, some including external 
shear, and conveyed the image position and time delay ratio information only 
to the other one of us, the modeler. In each case the modeler decided, based
on the appearance of the reconstructed galaxy mass maps whether inversion 
symmetry was appropriate or not. The modeler was told the correct answer for 
$h$, which is 0.025 in our synthetic universe, after the final $p(h)$ 
distributions were obtained. 

Distributions from the four synthetic lenses were multiplied together to 
yield the combined estimated $p(h)$ shown in Figure 1 as the solid line. 
The dashed histogram is for the case where inversion symmetry was 
imposed on all four galaxies, and the dotted histogram represents the case 
where inversion symmetry was not applied in any of the systems. All three 
resultant distributions recover $h$ fairly well, with the $90\%$ of the
models contained within $10\%$ of the true $h=0.025$. However the distributions
are not the same; the most probable values are different by $\sim 10\%$.
This illustrates how a relatively minor feature in modeling constraints, 
namely inclusion or exclusion of inversion symmetry, can make a considerable 
difference in the estimated $h$ value when the goal is to achieve precision 
of better than $10\%$. Based on this observation we conclude that the assumed 
galaxy shape in parametric reconstructions plays a major role in determining 
the outcome of \hub.

\section{PG1115+080 and B1608+656}

We now apply the method to PG1115 and B1608, both of which have accurate 
image position and time delay measurements (Schechter et al. 1997, 
Barkana 1997, Myers et al. 1995, Fassnacht et al. 1999). Figure 2 shows
an ensemble average of 100 reconstructed mass maps for PG1115. Note that the 
density contours are smooth and roughly elliptical.
\begin{figure}
\epsscale{0.5}
\plotone{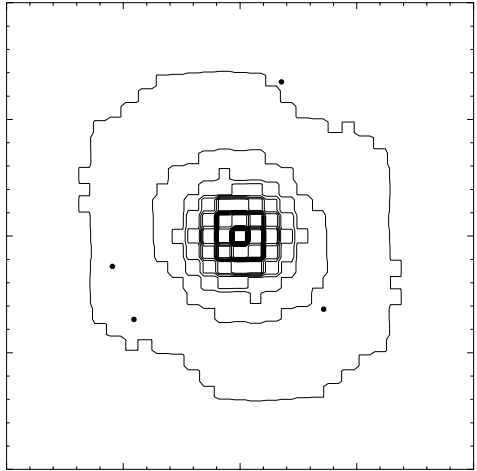}
\caption{An average of 100 reconstructed mass maps for PG1115. Contours are 
at $1\over 3$, $2\over 3$, 1, etc of $\Sigma_{crit}$. 
Images are also shown.}\label{fig-2}
\end{figure}
Figure 3 is a plot of the double logarithmic slope of the projected density
profile in the vicinity of the images vs. the derived estimate for \hub~ for
each of the 100 reconstructed galaxies. Since the anticorrelation is well 
defined and is understood in terms of the arrival time surface, it can
potentially be used as an additional modeling constraint.
\begin{figure}
\epsscale{0.33}
\plotone{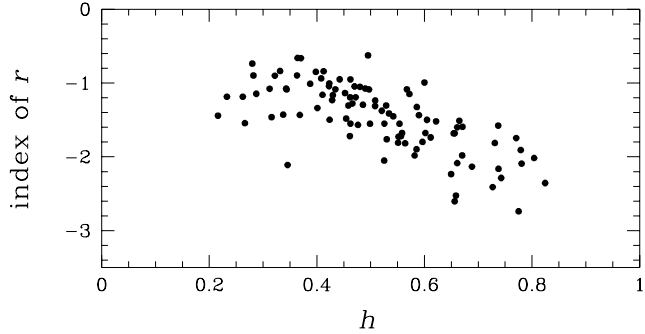}
\caption{The slope of the projected density profile in the image region vs. 
the derived \hub~ for each of the 100 reconstructed galaxies for the case of
PG1115. Isothermal slope is -1.}\label{fig-3}
\end{figure}
The combined $p(h)$ distribution based on the lensing data of PG1115 and B1608 
is presented in Figure 4. The median is at about 60 km~s$^{-1}$~Mpc$^{-1}$,
and the $90\%$ confidence range extends from 45 to 80 km~s$^{-1}$~Mpc$^{-1}$ .
\begin{figure}
\epsscale{0.33}
\plotone{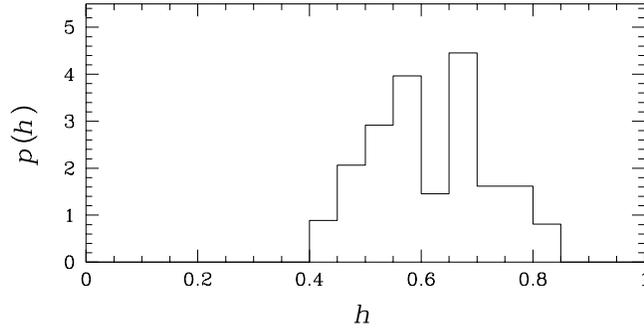}
\caption{Combined probability distribution based on the lensing data from
PG1115 and B1608.}\label{fig-4}
\end{figure}

\end{document}